\documentclass[pre,twocolumn,superscriptaddress]{revtex4} 

\usepackage{amsfonts}
\usepackage{amsmath}
\usepackage{amssymb}
\usepackage{booktabs}
\usepackage{caption}
\usepackage{color}
\usepackage{comment}
\usepackage{float}
\usepackage{graphicx}
\usepackage{hyperref}
\usepackage[utf8]{inputenc} 
\usepackage{subfig}
\usepackage{xspace}

\newcommand{\pygbe}{\texttt{PyGBe}\xspace}

\renewcommand{\O}[1]{\mathcal{O}(#1)}

\graphicspath{{figs/}} 

\begin{document}

\title{Computational nanoplasmonics in the quasistatic limit for biosensing applications}

\author{Natalia C. Clementi}
\email{ncclementi@gwu.edu}
\affiliation{Department of Mechanical \& Aerospace Engineering, The George Washington University, Washington, D.C.}

\author{Christopher D. Cooper}
\email{christopher.cooper@usm.cl}
\affiliation{Department of Mechanical Engineering and Centro Cient\'ifico Tecnol\'ogico de Valpara\'iso, Universidad T\'ecnica Federico Santa Mar\'ia, Valpara\'iso, Chile.}

\author{Lorena A.~Barba}
\email{labarba@gwu.edu}
\affiliation{Department of Mechanical \& Aerospace Engineering, The George Washington University, Washington, D.C.}

\begin{abstract} 

The phenomenon of localized surface plasmon resonance provides high sensitivity in detecting biomolecules through shifts in resonance frequency when a target is present. 
Computational studies in this field have used the full Maxwell equations with simplified models of a sensor-analyte system, or neglected the analyte altogether. 
In the long-wavelength limit, one can simplify the theory via an electrostatics approximation, while adding geometrical detail in the sensor and analytes (at moderate computational cost).
This work uses the latter approach, expanding the open-source \pygbe code to compute the extinction cross-section of metallic nanoparticles in the presence of any target for sensing.
The target molecule is represented by a surface mesh, based on its crystal structure. 
\pygbe is research software for continuum electrostatics, written in Python with computationally expensive parts accelerated on GPU hardware, via PyCUDA.
It is also accelerated algorithmically via a treecode that offers $\mathcal{O}(N \log N)$ computational complexity. 
These features allow \pygbe to handle problems with half a million boundary elements or more.
In this work, we demonstrate the suitability of \pygbe, extended to compute LSPR response in the electrostatic limit, for biosensing applications. 
Using a model problem consisting of an isolated silver nanosphere in an electric field, our results show grid convergence as $1/N$, and accurate computation of the extinction cross-section as a function of wavelength (compared with an analytical solution).
For a model of a sensor-analyte system, consisting of a spherical silver nanoparticle and a set of bovine serum albumin (BSA) proteins, our results again obtain grid convergence as $1/N$ (with respect to the Richardson extrapolated value).
Computing the LSPR response as a function of wavelength in the presence of BSA proteins captures a red-shift of 0.5 nm in the resonance frequency due to the presence of the analytes at 1-nm distance.
The final result is a sensitivity study of the biosensor model, obtaining the shift in resonance frequency for various distances between the proteins and the nanoparticle.
All results in this paper are fully reproducible, and we have deposited in archival data repositories all the materials needed to run the computations again and re-create the figures. \pygbe is open source under a permissive license and openly developed. Documentation is available at \url{http://barbagroup.github.io/pygbe/docs/}. 
\end{abstract}

\maketitle


\section{Introduction} \label{sec:intro}

\begin{figure}
 \centering
   \includegraphics[width=0.35\textwidth]{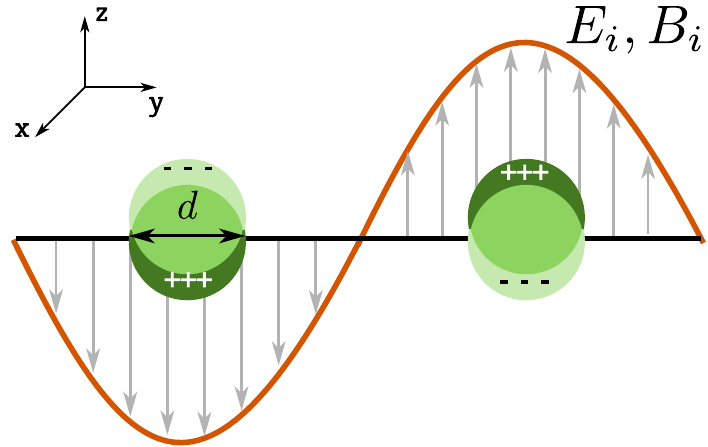} 
   \caption{Illustration of the localized surface plasmon resonance (LSPR) effect of a metallic nanoparticle under an electromagnetic field.    \label{fig:lspr}}
  \end{figure}

Localized surface plasmon resonance (LSPR) is an optical effect where an 
electromagnetic wave excites the free electrons on the surface of a metallic nanoparticle.
The vibrations of the electron cloud are known as plasmons, and in LSPR they resonate with the incoming
field (see Figure \ref{fig:lspr}). When this happens, most of the incoming energy is
either absorbed by the nanoparticle, or scattered in different directions, 
both effects creating a shadow behind the scatterer (a.k.a., extinction). 
In the case of nanoparticles smaller than 20 nm, absorption dominates and scattering 
contributions are negligible \cite{PetryayevaKrull2011, OlsonETal2015}. In LSPR, the 
wavelength of the incoming wave is often much larger than the size of the nanoparticle, 
which allows for valid approximations that simplify the mathematical model.

The phenomenon of LSPR can be used for biosensing, 
as the resonance frequency is highly dependent on the dielectric environment 
around the scatterer. 
The resonance frequency shifts whenever an analyte binds to the nanoparticle, 
resulting in a very sensitive means of detecting its presence \cite{HaesVanduyne2002,HaesETal2004}.

Numerical models for LSPR generally rely on the 
solution of Maxwell's equations in some form, using finite difference time-domain (FDTD),
boundary element, or finite element methods \cite{SolisTaboadaObelleiroLiz-MaarzanGarciadeabajo2014}. 
These methods have been used to study the 
optical properties of dielectric or metallic nanoparticles \cite{Hohenester2018,HohenesterTrugler2012,
JungPedersenSondergaardPedersenLarsenNielsen2010, VideenSun2003,
MayergoyzFredkinZhang2005, MayergoyzZhang2007}, interactions between nanoparticles
and electron beams \cite{GarciadeabajoAizpurua1997, GarciadeabajoHowie2002},
and surface plasmon resonance sensors.
In the latter application, researchers have used simple mathematical models for the 
interaction between a metallic nanoparticle and biomolecules,
like representing the medium and the dissolved analytes with an effective permittivity \cite{JungCampbellChinowskyMarYee1998,WilletsVandyune2007,PhanETal2013}, 
or representing the target molecules as spheres 
\cite{DavisGomezVernon2010,AntosiewiczApellClaudioKall2011}.

Progress in biosensor research is still predominantly made
through experimental investigations, which can often be costly and time consuming.
Computational approaches could assist the design process and play a role  
in optimizing biosensors, giving access to details that are not available in experimental settings.
For example, empirical studies showed that the sensitivity of the sensor
is highly dependent on the distance between the nanoparticle and the analyte \cite{HaesETal2004}.
These studies were complemented with models using a discrete dipole approximation (DDA),
which includes the effect of the analyte through the effective permittivity. 
Other experimental studies complemented by modeling fully ignore the presence of the target molecules.
For example, Beuwer et al.~\cite{BeuwervanHoofZijlstra2018} and Henkel et al.~\cite{HenkelETal2018} 
used a boundary element method (BEM) in studies of the sensitivity of plasmonic sensors 
relying on (at least) two metallic nanoparticles (one on the sensor and one attached to the analyte).
Explicitly including the target molecules in the model may be needed in some cases, however.
For instance, despite experimental evidence showing that LSPR sensors are sensitive enough to detect 
conformational changes of the analytes \cite{HallETal2011}, 
these simplified models are not able to capture such details.


Even though LSPR is an optical effect, electrostatic theory
provides a good approximation in the long-wavelength limit. This work uses
the boundary integral electrostatics solver \pygbe \cite{CooperETal2016} 
to compute the extinction cross-section of metallic nanoparticles, and to study how LSPR 
response changes in the presence of a biomolecule. 
We treat Maxwell's equations quasi-statically \cite{MayergoyzZhang2007} and
explicitly represent the target biomolecules by a surface mesh built from the crystal structures. 

\pygbe is a Python implementation of continuum electrostatic theory, used
for computing solvation energy of biomolecular systems. 
It has also been used to study protein orientation near charged nanosurfaces \cite{CooperClementiBarba2015}.
The code was recently extended to allow for complex dielectric constants 
\cite{ClementiETal2017}, aiming towards the LSPR biosensing application. 
The boundary element solver in \pygbe
is accelerated algorithmically via a treecode---an $\mathcal{O}(N\log N)$ fast-summation method---and on hardware by taking advantage of graphic processing units (GPUs). 
With these features, \pygbe is able to easily handle problems with in the order of 
half a million boundary elements, or more, 
allowing for the explicit representation of the biomolecular surface.
Other research software that could be used in this setting includes  
BEM++ \cite{SmigajETal2015} and a Matlab toolbox called MNPBEM \cite{HohenesterTrugler2012}, which have the capability to solve the full Maxwell's equations and the electrostatic approximation in the long-wavelength limit.
We believe in both cases the size of problems they can solve, in terms of number of boundary elements, may not be enough to resolve the details of target biomolecules from their crystal structure. 

The software is shared under the BSD 3-clause license 
and is openly developed via its repository on Github (\url{https://github.com/barbagroup/pygbe}).
This study also follows careful reproducibility practices, and all materials necessary
to reproduce the results are publicly available in reproducibility packages.
We use the Figshare and Zenodo services to deposit the computational meshes,
input and configuration files, and file bundles corresponding to the main figures in the paper.
See the figure captions for references to the open data artifacts.





\section{Methods}\label{sec:methods}

The original implementation of \pygbe used continuum electrostatic theory to compute
the solvation energy of biomolecular systems. In that setting, biomolecules are modeled as 
dielectric cavities inside an infinite continuum solvent, 
leading to a Poisson equation inside the molecules and Laplace or Poisson-Boltzmann in the solvent medium (with appropriate boundary conditions).
This set of partial differential equations can be 
expressed with the corresponding boundary integral equation along the molecular interface, 
which \pygbe solves using a boundary element method \cite{CooperBardhanBarba2013,CooperClementiBarba2015}.

The present work extends \pygbe to the LSPR biosensing application. 
In the long-wavelength limit, Maxwell's equations can be approximated by a Laplace equation,
which permits using the methods implemented in \pygbe, with modifications
to allow for complex-valued permittivities, and to include the
effect of an external electric field.
This section describes the mathematical formulation for computing electromagnetic scattering 
in the long-wavelength setting, and develops the associated boundary integral equations 
and their discretized form.

\subsection{Scattering of small particles} \label{sec:scattering_small}

\begin{figure}
   \centering
   \includegraphics[width=0.45\textwidth]{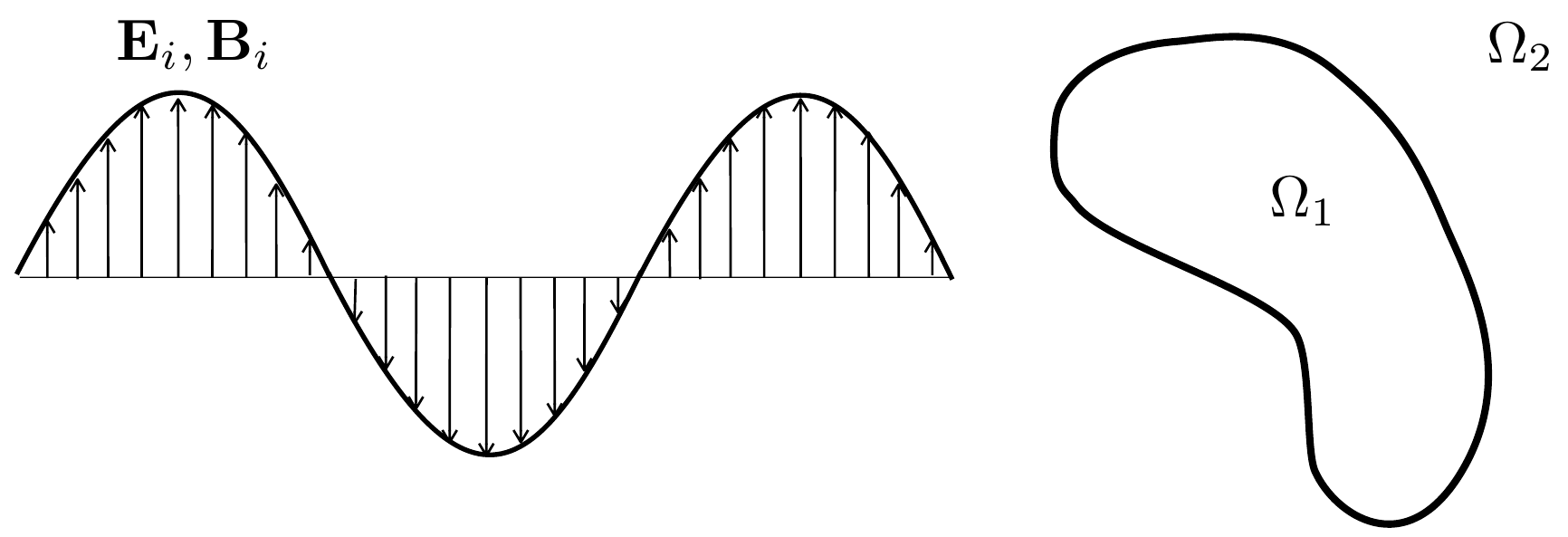} 
   \caption{Nanoparticle interacting with an electromagnetic wave.}
   \label{fig:part_wave}
\end{figure}

Electromagnetic scattering is usually modeled with Maxwell's equations.
When the wavelength of the incoming wave is much larger than the
scatterer, these can be reduced to a \emph{quasi-static} 
first-order approximation \cite{MayergoyzZhang2007}:
\begin{align} \label{eq:electrostatic_scatter_E}
\nabla \cdot \mathbf{E}_{1s} &= 0 \qquad \nabla \times \mathbf{E}_{1s} = 0, \nonumber \\
\nabla \cdot \mathbf{E}_{2s} &= 0 \qquad \nabla \times \mathbf{E}_{2s} = 0, \nonumber \\
\text{with interface conditions, } \nonumber \\
(\epsilon_1\mathbf{E}_{1s} - \epsilon_2\mathbf{E}_{2s})\cdot\mathbf{n} &= (\epsilon_2-\epsilon_1)\mathbf{E}_i\cdot \mathbf{n}.
\end{align}
In Equation \eqref{eq:electrostatic_scatter_E}, $\mathbf{E}_{1s}$ and $\mathbf{E}_{2s}$ 
are the electric fields of the scattered wave in the nanoparticle and host regions, respectively 
(see Figure \ref{fig:part_wave}), 
$\mathbf{E}_{i}$ is the field of the incoming wave, and $\epsilon_1$ 
and $\epsilon_2$ are the permittivities.
This approximation decouples the electric and magnetic fields, neglects the magnetic field, 
and describes the electric field as a curl-free vector field.
Hence, we can reformulate Equation \eqref{eq:electrostatic_scatter_E} with a scalar potential
($-\nabla \phi_{js} = \mathbf{E}_{js}$), as follows:
\begin{align} \label{eq:electrostatic_scatter}
\nabla^2 \phi_{1s} &= 0 \qquad \nabla^2 \phi_{2s} = 0 \qquad\text{on $\Omega_1$, $\Omega_2$} \nonumber \\
\epsilon_1\frac{\partial\phi_{1s}}{\partial \mathbf{n}} - \epsilon_2\frac{\partial\phi_{2s}}{\partial\mathbf{n}} &= (\epsilon_2-\epsilon_1)\frac{\partial\phi_i}{\partial\mathbf{n}} \quad \phi_{1s} = \phi_{2s} \quad \text{on $\Gamma$}.
\end{align}
Equation \eqref{eq:electrostatic_scatter} is an electrostatic equation 
with an imposed electric field $\mathbf{E}_i=-\nabla\phi_i$, where $\Gamma$ 
is the boundary between regions $\Omega_1$ and $\Omega_2$.

\subsection{Far-field scattering} \label{sec:ff_scattering}

In LSPR, the scattered electromagnetic wave is measured by a detector located far away 
from the scatterer (nanoparticle), and plasmon resonance is identified when the energy 
detected is minimum. In the far-field limit, the scattered field
in the outside region ($\Omega_2$) is given by: 

\begin{equation} \label{eq:scat_efield_long_range}
    \mathbf{E}_{2s} = \frac{1}{4\pi\epsilon_2}k^2\frac{e^{ikr}}{r} (\mathbf{\hat{r}} \times \mathbf{p})\times\mathbf{\hat{r}}.
\end{equation} 

\noindent where $k=2\pi/\lambda$ is the wave number and $\lambda$ the wavelength, $\mathbf{\hat{r}}$ 
is a unit vector in the direction of the observation point, and $\mathbf{p}$ is
the dipole moment.
We can obtain the scattered field using the 
scattering amplitude \cite{Jackson}:

\begin{equation} \label{eq:scat_efield_fwa}
    \mathbf{E}_{2s}(\mathbf{r})_{r\to\infty} = \frac{e^{ikr}}{r} \mathbf{F}(\mathbf{k},\mathbf{k}_0),
\end{equation}

\noindent where $\mathbf{F}$ is the scattering amplitude, $\mathbf{k}$ is the 
scattered wave vector in the direction of propagation, and $\mathbf{k}_0$ the 
wave vector of the incident field. 

\subsection{Extinction cross-section and optical theorem} \label{sec:cext_ot}

The extinction cross-section ($C_\text{ext}$) is a measure of the energy that 
does not reach the detector, either because of scattering in other directions,
or absorption. This quantity is defined as the ratio between the lost energy and 
the intensity of the incoming wave, and has units of area. 
The extinction cross-section peaks at resonance of plasmons.

The extinction cross-section is related to the forward-scattering amplitude via the optical theorem. 
The traditional expression for this relationship applies for non-absorbing media 
\cite{MayergoyzZhang2007, Jackson}; 
Mishchenko \cite{Mishchenko2007} corrected it for absorbing media, 
giving an expression that can be re-written using Jackson's notation \cite{Jackson} as follows:

\begin{equation} \label{eq:cext_fwa}
    C_\text{ext} = \frac{4\pi}{k^\prime} \operatorname{Im} \left[ \frac{\mathbf{\hat{e}}_i}{|\mathbf{E}_i|}\mathbf{F}(\mathbf{k}=\mathbf{k}_0, \mathbf{k}_0) \right].
\end{equation}

\noindent Here, $k^\prime$ is the real part of the complex wave number, 

\begin{equation}
    k = k^\prime + ik^{\prime\prime} = \frac{2\pi}{\lambda} n,
\end{equation}

\noindent and $n$ is the refraction index of the host medium.

Combining Equations \eqref{eq:scat_efield_long_range} and \eqref{eq:scat_efield_fwa},
we can compute the scattering amplitude to then obtain the extinction cross-section 
with Equation \eqref{eq:cext_fwa}.

\subsection{The boundary element method} \label{sec:lspr_bem}

\subsubsection{Electrostatic potential of a nanoparticle under an electric field} \label{sec:pot_elec_field}

\paragraph{Integral formulation}

Using Green's second identity, the system of partial differential equations 
in Equation \eqref{eq:electrostatic_scatter} can be rewritten as a system 
of boundary integral equations \cite{BrebbiaDominguez1992}. Evaluating on the surface $\Gamma$, this
becomes
\begin{align} \label{eq:integral_eq_lspr_nobc}
\frac{\phi_{1s,\Gamma}}{2}+ K_{L}^{\Gamma}(\phi_{1s,\Gamma}) - V_{L}^{\Gamma} \left(\frac{\partial}{\partial \mathbf{n}}\phi_{1s,\Gamma} \right) = 0&  \nonumber \\
\frac{\phi_{2s,\Gamma}}{2} - K_{L}^{\Gamma}(\phi_{2s,\Gamma}) + V_{L}^{\Gamma} \left( \frac{\partial}{\partial \mathbf{n}} \phi_{2s,\Gamma} \right) = 0&,
\end{align}
where $V$ and $K$ are the single- and double-layer operators, respectively:
\begin{equation}\label{eq:single_layer}
V^{\Gamma}_L (\psi(\mathbf{r}_\Gamma)) = \oint_\Gamma \psi(\mathbf{r}'_\Gamma) G_L(\mathbf{r}_\Gamma, \mathbf{r}'_\Gamma) \text{d} \Gamma',
\end{equation}
\begin{equation}\label{eq:double_layer}
K^{\Gamma}_L (\psi(\mathbf{r}_\Gamma)) = \oint_\Gamma \psi(\mathbf{r}'_\Gamma) \frac{\partial}{\partial \mathbf{n}}G_L(\mathbf{r}_\Gamma, \mathbf{r}'_\Gamma) \text{d} \Gamma'.
\end{equation}
\noindent Here, $G_L$ is the free-space Green's function of the Laplace equation:
\begin{equation}
G_L(\mathbf{r},\mathbf{r}') = \frac{1}{4\pi|\mathbf{r}-\mathbf{r}'|}
\end{equation}

\noindent Applying the interface conditions of Equation \eqref{eq:electrostatic_scatter},
leads to:
\begin{align} \label{eq:integral_eq_lspr}
\frac{\phi_{1s,\Gamma}}{2}+ K_{L}^{\Gamma}(\phi_{1s,\Gamma}) - V_{L}^{\Gamma} \left(\frac{\partial}{\partial \mathbf{n}}\phi_{1s,\Gamma} \right) &= 0  \nonumber \\
\frac{\phi_{1s,\Gamma}}{2} - K_{L}^{\Gamma}(\phi_{1s,\Gamma}) + \frac{\epsilon_1}{\epsilon_2}V_{L}^{\Gamma} \left( \frac{\partial}{\partial \mathbf{n}} \phi_{1s,\Gamma}  \right) &= \nonumber \\
 \frac{\epsilon_2-\epsilon_1}{\epsilon_2}V_{L}^{\Gamma}\left( \frac{\partial}{\partial \mathbf{n}} \phi_{i,\Gamma} \right)\quad \text{on $\Gamma$.}
\end{align}

\newpage
\subsubsection{Analyte-sensor electrostatic potential under an electric field}

\begin{figure}
    \centering
    \includegraphics[width=0.25\textwidth]{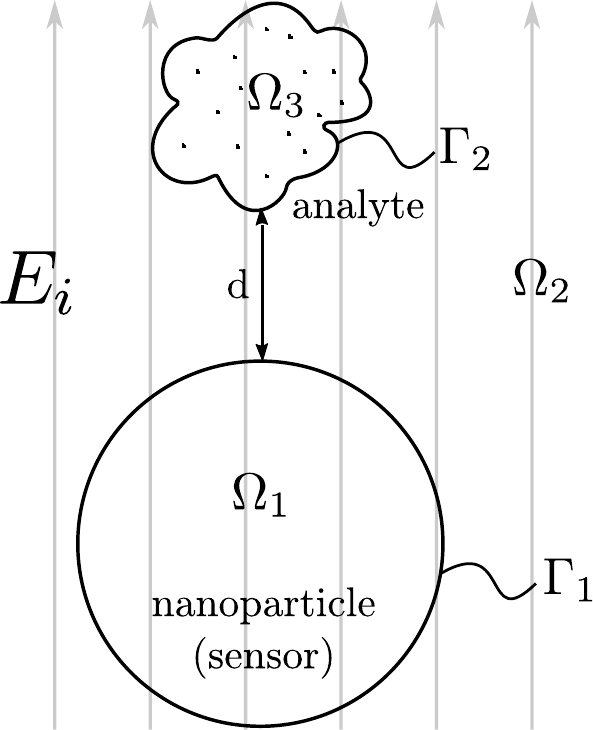} 
    \caption{Analyte-sensor system under electric field.}
    \label{fig:analyte-sensor}
 \end{figure}

The sketch in Figure \ref{fig:analyte-sensor} shows a metallic nanoparticle ($\Omega_1$) interacting with an analyte ($\Omega_3$), under an external electric field.
Mathematically, this situation can be modeled as

\begin{align}\label{eq:electrostatic_scatter_prot_sen}
\nabla^2 \phi_{1s} &= 0, \qquad \nabla^2 \phi_{2s} = 0 \qquad\text{on $\Omega_1$, $\Omega_2$} \nonumber\\
\nabla^2 \phi_{3s} &= -\frac{1}{\epsilon_3} \sum_{k=0}^{N_q} \delta(|\mathbf{r}-\mathbf{r}_k|) q_k \qquad\text{on $\Omega_3$} \nonumber \\
\epsilon_1\frac{\partial\phi_{1s}}{\partial \mathbf{n}} - \epsilon_2\frac{\partial\phi_{2s}}{\partial\mathbf{n}} &= (\epsilon_2-\epsilon_1)\frac{\partial\phi_i}{\partial\mathbf{n}} \quad \phi_{1s} = \phi_{2s} \quad \text{on $\Gamma_1$}. \nonumber\\
\epsilon_3\frac{\partial\phi_{3s}}{\partial \mathbf{n}} - \epsilon_2\frac{\partial\phi_{2s}}{\partial\mathbf{n}} &= (\epsilon_2-\epsilon_3)\frac{\partial\phi_i}{\partial\mathbf{n}} \quad \phi_{3s} = \phi_{2s} \quad \text{on $\Gamma_2$}.
\end{align}
where $q_k$ are the point charges of the atoms inside the protein, located at $\mathbf{r}_k$.

\paragraph{Integral formulation}

Similar to Equation \eqref{eq:integral_eq_lspr}, we can write the system of partial differential equations in \eqref{eq:electrostatic_scatter_prot_sen} as

\begin{widetext} 

\begin{align} \label{eq:integral_eq_lspr_nobc_system}
\frac{\phi_{1s,\Gamma_1}}{2}+ K_{L,\Gamma_1}^{\Gamma_1}(\phi_{1s,\Gamma_1}) - V_{L,\Gamma_1}^{\Gamma_1} \left(\frac{\partial}{\partial \mathbf{n}}\phi_{1s,\Gamma_1} \right) &= 0  \nonumber \\
\frac{\phi_{2s,\Gamma_1}}{2} - K_{L,\Gamma_1}^{\Gamma_1}(\phi_{2s,\Gamma_1}) + V_{L,\Gamma_1}^{\Gamma_1} \left(\frac{\partial}{\partial \mathbf{n}}\phi_{2s,\Gamma_1} \right) 
 - K_{L,\Gamma_2}^{\Gamma_1}(\phi_{2s,\Gamma_2}) + V_{L,\Gamma_2}^{\Gamma_1} \left(\frac{\partial}{\partial \mathbf{n}}\phi_{2s,\Gamma_2} \right) &= 0  \nonumber \\
\frac{\phi_{2s,\Gamma_2}}{2} - K_{L,\Gamma_1}^{\Gamma_2}(\phi_{2s,\Gamma_1}) + V_{L,\Gamma_1}^{\Gamma_2} \left(\frac{\partial}{\partial \mathbf{n}}\phi_{2s,\Gamma_1} \right)  
- K_{L,\Gamma_2}^{\Gamma_2}(\phi_{2s,\Gamma_2}) + V_{L,\Gamma_2}^{\Gamma_2} \left(\frac{\partial}{\partial \mathbf{n}}\phi_{2s,\Gamma_2} \right) &= 0  \nonumber \\
\frac{\phi_{3s,\Gamma_2}}{2} + K_{L,\Gamma_2}^{\Gamma_2}(\phi_{3s,\Gamma_2}) - V_{L,\Gamma_2}^{\Gamma_2} \left( \frac{\partial}{\partial \mathbf{n}} \phi_{3s,\Gamma_2} \right) &= \frac{1}{4\pi\epsilon_3} \sum_{k=0}^{N_q} \frac{q_k}{|\mathbf{r}_{\Gamma_2} - \mathbf{r}_k|} ,
\end{align}
\noindent where $V$ and $K$ are the single- and double-layer operators in equations 
\eqref{eq:single_layer} and \eqref{eq:double_layer}. In this case, we distinguish between the
surface where the integrals run (subindex), and the surface that contains the evaluation point (superindex).

Applying the interface conditions of equation \eqref{eq:electrostatic_scatter_prot_sen},
leads to:

\begin{align} \label{eq:integral_eq_lspr_system}
\frac{\phi_{1s,\Gamma_1}}{2}&+ K_{L,\Gamma_1}^{\Gamma_1}(\phi_{1s,\Gamma_1}) - V_{L,\Gamma_1}^{\Gamma_1} \left(\frac{\partial}{\partial \mathbf{n}}\phi_{1s,\Gamma_1} \right) = 0  \nonumber \\
 \frac{\phi_{1s,\Gamma_1}}{2}& - K_{L,\Gamma_1}^{\Gamma_1}(\phi_{1s,\Gamma_1}) + V_{L,\Gamma_1}^{\Gamma_1} \left(\frac{\epsilon_1}{\epsilon_2}\frac{\partial}{\partial \mathbf{n}}\phi_{1s,\Gamma_1} \right) - V_{L,\Gamma_1}^{\Gamma_1} \left(\frac{\epsilon_2-\epsilon_1}{\epsilon_2}\frac{\partial}{\partial \mathbf{n}}\phi_{i,\Gamma_1} \right) \nonumber\\ 
 & - K_{L,\Gamma_2}^{\Gamma_1}(\phi_{3s,\Gamma_2}) + V_{L,\Gamma_2}^{\Gamma_1} \left(\frac{\epsilon_3}{\epsilon_2}\frac{\partial}{\partial \mathbf{n}}\phi_{3s,\Gamma_2} \right)  - V_{L,\Gamma_2}^{\Gamma_1} \left(\frac{\epsilon_2 -\epsilon_3}{\epsilon_2}\frac{\partial}{\partial \mathbf{n}}\phi_{i,\Gamma_2} \right) = 0   \nonumber \\
 \frac{\phi_{3s,\Gamma_1}}{2}& - K_{L,\Gamma_1}^{\Gamma_2}(\phi_{1s,\Gamma_1}) + V_{L,\Gamma_1}^{\Gamma_2} \left(\frac{\epsilon_1}{\epsilon_2}\frac{\partial}{\partial \mathbf{n}}\phi_{1s,\Gamma_1} \right) - V_{L,\Gamma_1}^{\Gamma_2} \left(\frac{\epsilon_2-\epsilon_1}{\epsilon_2}\frac{\partial}{\partial \mathbf{n}}\phi_{i,\Gamma_1} \right) \nonumber \\
& - K_{L,\Gamma_2}^{\Gamma_2}(\phi_{3s,\Gamma_2}) + V_{L,\Gamma_2}^{\Gamma_2} \left(\frac{\epsilon_3}{\epsilon_2}\frac{\partial}{\partial \mathbf{n}}\phi_{3s,\Gamma_2} \right)  - V_{L,\Gamma_2}^{\Gamma_2} \left(\frac{\epsilon_2 -\epsilon_3}{\epsilon_2}\frac{\partial}{\partial \mathbf{n}}\phi_{i,\Gamma_2} \right) = 0  \nonumber \\
\frac{\phi_{3s,\Gamma_2}}{2}& + K_{L,\Gamma_2}^{\Gamma_2}(\phi_{3s,\Gamma_2}) - V_{L,\Gamma_2}^{\Gamma_2} \left( \frac{\partial}{\partial \mathbf{n}} \phi_{3s,\Gamma_2} \right) = \frac{1}{4\pi\epsilon_3} \sum_{k=0}^{N_q} \frac{q_k}{|\mathbf{r}_{\Gamma_2} - \mathbf{r}_k|} 
\end{align}
\end{widetext}

\paragraph{Discretization and linear system}

We discretize the surface into flat triangles, and assume that  $\phi$ and 
$\partial \phi/\partial \mathbf{n}$ are constant within each element. We can
then write the layer operators in their discretized form as follows:
\begin{align} \label{eq:layers_disc}
V_{L,\text{disc}}^{\mathbf{r}_\Gamma} \left( \frac{\partial}{\partial \mathbf{n}} \phi(\mathbf{r}_{\Gamma}) \right) &= \sum_{j=1}^{N_p} \frac{\partial}{\partial \mathbf{n}} \phi(\mathbf{r}_{\Gamma_j}) \int_{\Gamma_j} G_L(\mathbf{r}_\Gamma,\mathbf{r}_{\Gamma_j})  \mathrm{d} \Gamma_j  \nonumber \\
K_{L,\text{disc}}^{\mathbf{r}_\Gamma}(\phi(\mathbf{r}_{\Gamma})) &=  \sum_{j=1}^{N_p}\phi(\mathbf{r}_{\Gamma_j})\int_{\Gamma_j} \frac{\partial}{\partial \mathbf{n}} \left[ G_L(\mathbf{r}_\Gamma,\mathbf{r}_{\Gamma_j}) \right]\mathrm{d} \Gamma_j
\end{align}
\noindent where $N_p$ is the number of discretization elements on $\Gamma$, 
and $\phi(\mathbf{r}_{\Gamma_j})$ and $\frac{\partial}{\partial \mathbf{n}} 
\phi(\mathbf{r}_{\Gamma_j})$ are the values of $\phi$ and 
$\frac{\partial \phi}{\partial \mathbf{n}}$ on panel $\Gamma_j$.
Using centroid collocation, we can write equation \eqref{eq:integral_eq_lspr} in matrix form as:
 \begin{equation} \label{eq:matrix_lspr}
 \left[
    \begin{matrix} 
       \frac{1}{2} + K_{L}^{\Gamma} & -V_{L}^{\Gamma}  \vspace{0.2cm} \\
       \frac{1}{2} - K_{L}^{\Gamma} &  \frac{\epsilon_1}{\epsilon_2} V_{L}^{\Gamma}  \vspace{0.2cm} 
    \end{matrix}
    \right] \left[ 
    \begin{matrix} 
       \phi_{1s,\Gamma} \vspace{0.2cm} \\
       \frac{\partial}{\partial \mathbf{n}} \phi_{1s,\Gamma} \vspace{0.2cm}
    \end{matrix} 
     \right] =   
    \left[
    \begin{matrix} 
       0 \\
       V_{L}^{\Gamma} \left(\frac{\epsilon_2-\epsilon_1}{\epsilon_2} \frac{\partial\phi_i}{\partial\mathbf{n}}\right) \vspace{0.2cm} 
    \end{matrix}
    \right]
 \end{equation}
Equation \eqref{eq:integral_eq_lspr_system} can be represented as:
\begin{align} \label{eq:matrix_multi}
 \left[
    \begin{matrix} 
       \frac{1}{2}+K_{L, \Gamma_1}^{\Gamma_1} & -V_{L, \Gamma_1}^{\Gamma_1} & 0 &  0   \vspace{0.2cm} \\
       \frac{1}{2}-K_{L, \Gamma_1}^{\Gamma_1} & \frac{\epsilon_1}{\epsilon_2} V_{L, \Gamma_1}^{\Gamma_1} & -K_{L, \Gamma_2}^{\Gamma_1} & \frac{\epsilon_3}{\epsilon_2} V_{L, \Gamma_2}^{\Gamma_1} \vspace{0.2cm}  \\
        -K_{L, \Gamma_1}^{\Gamma_2}&\frac{\epsilon_1}{\epsilon_2} V_{L, \Gamma_1}^{\Gamma_2} & \frac{1}{2}-K_{L, \Gamma_2}^{\Gamma_2}  &  \frac{\epsilon_3}{\epsilon_2} V_{L, \Gamma_2}^{\Gamma_2} \vspace{0.2cm} \\
       0 & 0 & \frac{1}{2}+K_{L, \Gamma_2}^{\Gamma_2}&  - V_{L, \Gamma_2}^{\Gamma_2}   \vspace{0.2cm} \\
    \end{matrix}
    \right] 
\cdot
 \left[
    \begin{matrix}
    \phi_{1,\Gamma_1} \vspace{0.2cm} \\
    \frac{\partial}{\partial \mathbf{n}} \phi_{1,\Gamma_1} \vspace{0.2cm} \\
    \phi_{3,\Gamma_2} \vspace{0.2cm} \\
    \frac{\partial}{\partial \mathbf{n}} \phi_{3,\Gamma_2} \vspace{0.2cm} \\
    \end{matrix}
\right]&
 \nonumber \\
 = \left[
    \begin{matrix}
    0 \vspace{0.2cm} \\
    V_{L,\Gamma_1}^{\Gamma_1} \left(\frac{\epsilon_2-\epsilon_1}{\epsilon_2}\frac{\partial}{\partial \mathbf{n}}\phi_{i,\Gamma_1} \right)
    + V_{L,\Gamma_2}^{\Gamma_1} \left(\frac{\epsilon_2 -\epsilon_3}{\epsilon_2}\frac{\partial}{\partial \mathbf{n}}\phi_{i,\Gamma_2} \right)
    \vspace{0.2cm}\\
    V_{L,\Gamma_1}^{\Gamma_2} \left(\frac{\epsilon_2-\epsilon_1}{\epsilon_2}\frac{\partial}{\partial \mathbf{n}}\phi_{i,\Gamma_1} \right)
    + V_{L,\Gamma_2}^{\Gamma_2} \left(\frac{\epsilon_2 -\epsilon_3}{\epsilon_2}\frac{\partial}{\partial \mathbf{n}}\phi_{i,\Gamma_2} \right)
    \vspace{0.2cm}\\
    \frac{1}{4\pi\epsilon_3}\sum_{k=0}^{N_q} \frac{q_k}{|\mathbf{r}_{\Gamma_2} - \mathbf{r}_k|} \vspace{0.2cm}  \\
    \end{matrix}
\right]&
\end{align}
\noindent where the elements of the matrix are
\begin{align} \label{eq:layers_element}
V_{L,ij}^{\Gamma} &= \int_{\Gamma_j} G_L(\mathbf{r}_{\Gamma_i},\mathbf{r}_{\Gamma_j})  \mathrm{d} \Gamma_j, \nonumber \\
K_{L,ij}^{\Gamma} &= \int_{\Gamma_j} \frac{\partial}{\partial \mathbf{n}} \left[ G_L(\mathbf{r}_{\Gamma_i},\mathbf{r}_{\Gamma_j}) \right]\mathrm{d} \Gamma_j,
\end{align}
\noindent with $\mathbf{r}_{\Gamma_i}$ being at the center of panel $\Gamma_i$.

\paragraph{Integral evaluation}

We evaluate the integrals in Equation \eqref{eq:layers_element} with Gauss quadrature
rules. The $1/r$ singularity of the Green's function poses a
problem to obtaining good accuracy when the integral is 
singular or near-singular. Therefore, we define three different regions, as follows.
\begin{description}
\item[Singular integrals:] If the collocation point is in the integration element,
the singularity is difficult to resolve with standard
Gauss integration schemes. In this case, we use a semi-analytical technique 
\cite{HessSmith1967,ZhuHuangSongWhite2001} that places $N_k$ quadrature nodes on the 
edges of the triangle.

\item[Near-singular integrals:] If the collocation point is close to the integration element,
the integrand has a high gradient, and high-order quadrature rules are required. 
We use the representative length of the integrated triangle ($L = \sqrt{2\cdot\text{Area}}$)
to define a threshold of the \emph{nearby} region, for example, when the integration panel 
is $2L$ or less away from the collocation point. For near-singular integrals, we use 
$K_{fine}=19, 25  \text{ or }  37$ points per triangle. 

\item[Far-away integrals:] When the distance between the collocation point and the integration
element is beyond the threshold, they are considered to be far-away. 
At this point, the integrand is smooth enough that we obtain good 
accuracy with low-order integration, for example, with 
$K=1, 3  \text{ or } 4$ Gauss quadrature points per boundary element. 
\end{description}

\subsubsection{Boundary integral expression of the dipole moment}

As shown in Equation \eqref{eq:scat_efield_long_range}, the scattered electric 
field in the far-away limit depends on the dipole moment. The dipole moment is 
defined as 
\begin{equation} \label{eq:dipole_def}
\mathbf{p} = \int_\Omega \mathbf{r} \rho \text{d}\Omega,
\end{equation}
and rewriting this equation using Gauss' law, we obtain
\begin{equation} \label{eq:dipole_def_gauss}
\mathbf{p} = -\epsilon_2\int_\Omega \mathbf{r} \nabla^2 \phi_{2s} \text{d}\Omega.
\end{equation}
For component $i$, this becomes:
\begin{equation} \label{eq:dipole_def_gauss_i}
{p_i} = -\epsilon_2\int_\Omega {x_i} \nabla^2 \phi_{2s} \text{d}\Omega.
\end{equation}
Using the identity
\begin{equation} \label{eq:identity_grad}
  \nabla \cdot \left(f \mathbf{v}\right) = \left( \nabla f \right)\cdot \mathbf{v} + f\left(\nabla \cdot \mathbf{v}\right)
\end{equation}
with $f=x_i$ and $\mathbf{v} = \nabla\phi_{2s}$, we can rewrite Equation \eqref{eq:dipole_def_gauss_i}
as 
\begin{equation}
- \frac{p_i}{\epsilon_2} = \int_\Omega \nabla \cdot \left( x_i \nabla \phi_{2s} \right) \; \text{d}\Omega - \int_\Omega \nabla x_i \cdot \nabla\phi_{2s} \; \text{d}\Omega, \nonumber 
\end{equation}
\noindent and applying the divergence theorem
\begin{equation} \label{eq:dip_gauss_interm_1}
- \frac{p_i}{\epsilon_2}= \oint_\Gamma  x_i  \nabla \phi_{2s} \cdot \mathbf{n} \; \text{d}\Gamma - \int_\Omega \nabla x_i \cdot \nabla\phi_{2s} \; \text{d}\Omega.
\end{equation}
Using the identity \eqref{eq:identity_grad} again in Equation \eqref{eq:dip_gauss_interm_1}, this time 
taking $f=\phi_{2s}$ and $\mathbf{v} = \nabla x_i$, we get:
\begin{align} \label{eq:dip_gauss_interm_2}
 - \frac{p_i}{\epsilon_2} =& \oint_\Gamma  x_i  \frac{\partial \phi_{2s}}{\partial \mathbf{n}} \text{d}\Gamma - \nonumber \\
 & \left[ \int_\Omega \nabla \cdot \left( \phi_{2s} \nabla x_i \right)\;\text{d}\Omega - \int_\Omega  \phi_{2s} \nabla^2 x_i \;\text{d}\Omega\right] \nonumber\\
=& \oint_\Gamma  x_i  \frac{\partial \phi_{2s}}{\partial \mathbf{n}} \; \text{d}\Gamma - \oint_\Gamma \phi_{2s} \nabla x_i \cdot \mathbf{n} \; \text{d}\Gamma \nonumber \\
=& \oint_\Gamma  x_i  \frac{\partial \phi_{2s}}{\partial \mathbf{n}} \; \text{d}\Gamma - \oint_\Gamma \phi_{2s} n_i \;\text{d}\Gamma
\end{align}
Throughout this derivation, the normals are pointing into $\Omega_1$. However, in our implementation 
all normals are pointing outwards, and we need to include an extra negative sign, yielding:
\begin{equation} \label{eq:dipole_def_gauss_i_final}
{p_i} = \epsilon_2 \left[ \oint_\Gamma  x_i  \frac{\partial \phi_{2s}}{\partial \mathbf{n}} \text{d}\Gamma - \oint_\Gamma \phi_{2s} n_i \; \text{d}\Gamma \right].
\end{equation}

Using BEM, we obtain the electrostatic potential and its normal derivative, on the surface of the nanoparticle, 
which we use in Equation \eqref{eq:dipole_def_gauss_i_final} to get the dipole 
moment, and in Equation \eqref{eq:scat_efield_long_range} to obtain the scattered
electric field. We can then use Equation \eqref{eq:scat_efield_fwa} and Equation 
\eqref{eq:cext_fwa} to get the extinction cross section.

\subsection{Acceleration strategies} \label{sec:acc_strategies}

One disadvantage of the Boundary Element Method (BEM) is that it generates dense matrices
after discretization. Solving the resulting linear system using
Gaussian elimination would require $\O{N^3}$ computations and $\O{N^2}$ storage, whereas for a
Krylov-subspace iterative solver, like the Generalized Minimal Residual Method (GMRES),
computations drop to $\O{N^2}$ because they are dominated by dense matrix-vector 
products. This makes BEM inefficient with more than a few thousand boundary elements,
which are the mesh sizes required for real applications. 

In our formulation with Gaussian quadrature and collocation, the matrix-vector product
becomes an $N$-body problem, with Gauss nodes acting as centers of mass (\emph{sources}), 
and the collocation points acting as evaluation points for the potential (\emph{targets}).
To overcome the unfavorable scaling,
we accelerate the matrix-vector product using a treecode algorithm \cite{BarnesHut1986,DuanKrasny2001}, 
which is a fast-summation algorithm capable of reducing $\O{N^2}$
computational patterns like
\begin{equation} \label{eq:summation}
V(\mathbf{x}_i) = \sum_{j=1}^{N} q_j \psi(\mathbf{x}_i, \mathbf{y}_j) 
\end{equation}
\noindent to a computational complexity of $\O{N \log N}$. In Equation \eqref{eq:summation} 
$q_j$ is the weight, $\psi$ the kernel, $\mathbf{y}_j$ the locations of sources and 
$\mathbf{x}_i$ the locations of targets.

The treecode groups sources geometrically in boxes of an octree, built ensuring
that no box in the lowest level has more than $N_\text{crit}$ sources. If a group of
sources is far away from a target, their influence is aggregated at an expansion center,
and the target interacts with the box, rather than with each source independently.
If the group of targets is  close, the treecode queries the child
boxes. If the box has no children and still is not far enough, the interaction is 
performed directly via \eqref{eq:summation}.
 The threshold to decide if a box is far enough is called the multipole-
acceptance criterion (MAC), defined as:
\begin{equation}
\theta > \frac{r_b}{r},
\end{equation}
\noindent where $r_b$ is the box size and $r$ the distance between the box center and the target.
Common values of $\theta$ are $1/2$ and $2/3$.
To approximate the contribution of the sources, we use Taylor expansions
of order $P$.
The treecode allows us to control the accuracy of the approximation by modifying $\theta$ and $P$.
Further details of the treecode implementation in \pygbe can be found in \cite{CooperBarba-share154331,CooperBardhanBarba2013}.

\subsection{Code modifications and added features} \label{sec:code_imp}

As mentioned at the beginning of this section, the present work extends the \pygbe code
to allow its application to nano-plasmonics. 
The code required the following modifications and added features:

\begin{itemize}
    \item Re-writing the GMRES solver to accept complex numbers. 
    \item Splitting treecode calculations into real and imaginary parts.
    \item Re-formatting configuration files to include electric field intensity and  wavelength.
    \item Adding the new function \texttt{read\_electric\_field}, to read the electric field intensity and its wavelength from configuration files.
    \item Adding the new function \texttt{dipole\_moment} to compute numerically the dipole moment by Equation \eqref{eq:dipole_def_gauss_i_final}.
    \item Adding a new function to compute the  extinction cross section (\texttt{extinction\_cross\_section}).
    \item Organizing LSPR computations on a different main script (called \texttt{lspr.py}).
\end{itemize}

\noindent For information about how to use the code, run examples and tests, see the
\pygbe documentation at \url{http://barbagroup.github.io/pygbe/docs/}

\subsection{Protein mesh preparation}
In Figure \ref{fig:analyte-sensor}, $\Omega_3$ is a region that represents the analyte molecule, which contains a point charge distribution of the partial charges, and is interfaced with the solvent by $\Gamma_2$, the solvent excluded surface (\texttt{SES}).
The \texttt{SES} is generated by rolling a spherical probe of the size of a water molecule ($1.4$\AA~ radius) around the analyte, and tracking the points where the probe and molecule make contact.
The open-source software Nanoshaper \cite{Nanoshaper} uses the molecular structure to produce a triangulation of the \texttt{SES}, which can be read by our software.
In particular, Nanoshaper takes as inputs the atomic coordinates, obtained from the Protein Data Bank, and radii, which were 
extracted from a \texttt{pqr} file generated with \texttt{pdb2pqr} \cite{Dolinsky04}.
We obtained the charge and van der Waals parameters of the analyte from \texttt{pdb2pqr} using the built-in \texttt{amber} force field.
In support of the reproducibility of our results, we deposited the final meshes in the Zenodo data repository.
See section \ref{sec:repro} for details.

\section{Results} \label{sec:results}

We present results for two kinds of problems. 
The first is a model problem for which an analytical solution is available, 
allowing for a grid-refinement study and code verification using that solution.
It consists of a spherical nanoparticle in a constant electric field, 
where the extinction cross-section can be derived in closed form.
The second set of results use a model for a biosensor detecting a target molecule, 
via frequency shifts in the plasmon resonance of a metallic nanoparticle. 
In this case, since an analytical solution is not available, we can use Richardson 
extrapolation to estimate the errors in a grid-refinement study.
We also computed the variation of the extinction cross-section with respect to wavelength
for the isolated nanoparticle, and in the presence of bovine serum albumin (BSA) proteins, 
varying the location of the analytes.
The final result is a sensitivity study of the biosensor model, 
looking at how the peak in frequency response varies with distance of the protein 
to the nanoparticle.

All results were obtained on a lab workstation, built from parts.
Hardware specifications are as follows: 
\begin{itemize}
  \item CPU: Intel Core i7-5930K Haswell-E 6-Core 3.5GHz LGA 2011-v3
  \item RAM: G.SKILL Ripjaws 4 series 32GB (4 x 8GB)
  \item GPU: Nvidia Tesla K40c (with 12 GB memory)
\end{itemize}

\subsection{Grid convergence and verification with an isolated silver nanoparticle} \label{sec:verification}

\begin{figure}[h] 
   \centering
   \includegraphics[width=0.3\textwidth]{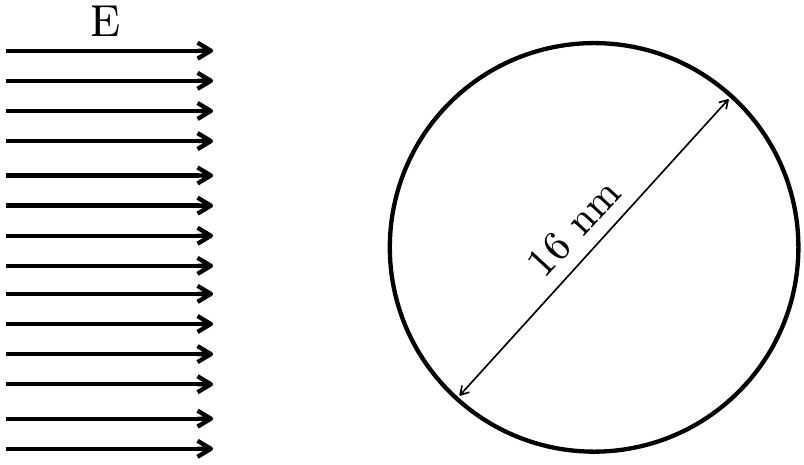} 
   \caption{Spherical nanoparticle in a constant electric field.}
   \label{fig:np_elec_field}
\end{figure}

\noindent In the long-wavelength limit, the electrostatic approximation applies and
the electromagnetic scattering of a small spherical particle can be modeled
by a sphere in a constant electric field. 
Figure \ref{fig:np_elec_field} illustrates this scenario.

This model problem has an analytical solution, which allows us to compare with
the numerical calculations of the extinction cross-section obtained with \pygbe,
for code verification and grid-convergence analysis.
Mishchenko \cite{Mishchenko2007} derived the following analytical result, 
valid for lossy mediums:
\begin{equation} 
    C_\text{ext} = \frac{4\pi a^3}{k^\prime} \operatorname{Im}\left(k^2 
                    \frac{\epsilon_p/\epsilon_m -1}{\epsilon_p/\epsilon_m +2}\right).
    \label{eq:an_sol}
\end{equation}
Here, $a$ is the radius of the sphere, $k$ is the complex wave number ($k=k^\prime +i k^{\prime\prime}$), $\epsilon_p$ 
is the dielectric constant of the particle, and $\epsilon_m$ is the dielectric constant
of the host medium. If the medium is not lossy, then $k^{\prime\prime}=0$ and $k=k^\prime$.

We completed a grid-convergence study of \pygbe for the extinction
cross-section of a spherical silver nanoparticle of radius 8 nm immersed in water,
under a $z$-polarized electric field with a wavelength of 380 nm and intensity of 
$-0.0037 e/({\AA}^2 \, \epsilon_0)$. In these conditions, water has a dielectric
constant of $1.7972 \, + \, 8.5048^{-09}i$ \cite{JohnsonChristy1972} and silver of
$-3.3877 \, + \, 0.1922i$ \cite{HaleQuerry1972}. 
Table \ref{table:quadparams1} lists the Gauss quadrature points used for each type of boundary element. 
The threshold parameter defining the near-singular region was 0.5 
(refer to the \pygbe documentation, under ``Parameter file format'').
Table \ref{table:treeparams1} shows the treecode and solver parameters for this grid-convergence study.

\begin{table}[h]
    \centering
    \caption{\label{table:quadparams1} Grid-convergence study: Gauss quadrature points; 
    $K$ and $K_{fine}$ are per element; $N_k $ is per element edge (semi-analytical integration). } 
    \begin{tabular}{l l}
    \hline
     distant elements: & $K=4$ \\
     near-singular integrals:   & $ K_{fine}=37$ \\
     singular elements:  & $N_k =9$ \\
    \hline
    \end{tabular}
\end{table}

\begin{table}[h]
    \centering
    \caption{\label{table:treeparams1} Grid-convergence study: treecode and solver parameters.} 
    \begin{tabular}{l l}
    \hline
    treecode order of expansion: & $P=15$\\
    MAC                                         & $\theta=0.5$\\
    GMRES tolerance                    & $10^{-5}$\\
    \hline
    \end{tabular}
\end{table}

\begin{figure}[t] 
   \centering
   \includegraphics[width=0.45\textwidth]{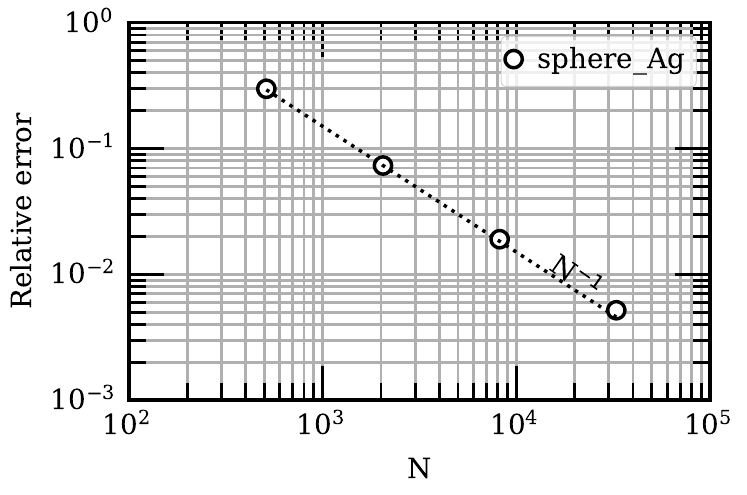} 
   \caption{Grid-convergence study for the extinction cross-section of a spherical silver
            nanoparticle, computed with \pygbe. Figure, plotting script and auxiliary files available under \textsc{cc-by} \cite{ClementiETal2018c}.}
   \label{fig:error_sphere_Ag}
\end{figure}

The results are shown in Figure \ref{fig:error_sphere_Ag}, where the mesh sizes are
512, 2048, 8192, and 32768 elements. 
The analytical solution with equation \eqref{eq:an_sol} is $C_{ext} = 1854.48$ nm$^2$, 
and the computed errors are as shown in Table \ref{table:err_iso_sphere}.
The dashed line in Figure \ref{fig:error_sphere_Ag} shows a $1/N$ slope, 
and the observed order of convergence is $0.98$,  
evidence that the meshes are correctly resolving the numerical solutions with \pygbe.

\begin{table}
    \centering
    \caption{\label{table:err_iso_sphere} Percentage error in the grid-convergence cases with an isolated silver nanosphere.} 
    \begin{tabular}{c c}
    \hline
    N & \% error \\
    \hline
     $512$ & $29.86$ \\
     $2048$ & $7.33$ \\
     $8192$ & $1.9$ \\
     $32768$ & $0.52$ \\
    \hline
    \end{tabular}
\end{table}

\begin{figure}[h] 
   \centering
   \includegraphics[width=0.45\textwidth]{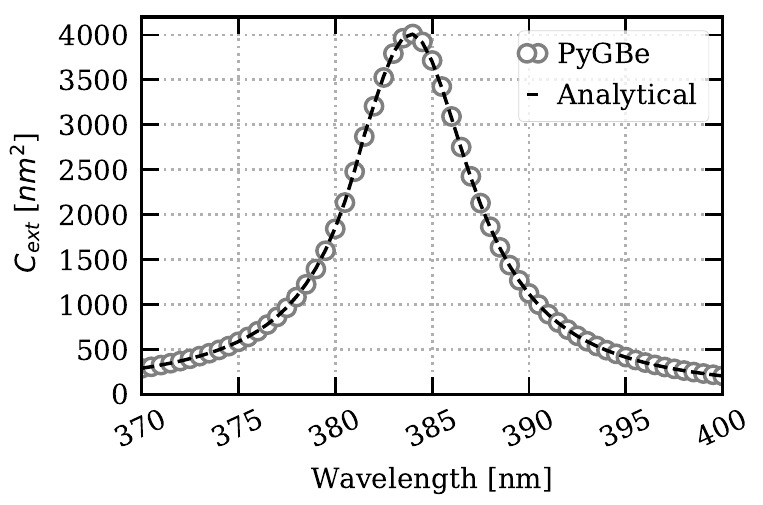} 
   \caption{Extinction cross-section as a function of wavelength for an $8$-nm
            silver sphere immersed in water. The peak in the values of extinction cross-section corresponds to the plasmon resonance of the metallic nanoparticle under the incoming electric field. Figure, plotting script and auxiliary files available under \textsc{cc-by} \cite{ClementiETal2018d}.}
   \label{fig:verif_sphere}
\end{figure}

As another verification test of \pygbe in the LSPR setting, we computed the extinction cross-section of an 
isolated sphere for a range of wavelengths. 
The results are shown in Figure \ref{fig:verif_sphere}, comparing with the analytical solution. 
The values of the dielectric constant for each wavelength were obtained by interpolation of 
experimental data \cite{JohnsonChristy1972, HaleQuerry1972}.
For reproducibility of these results, we provide a Jupyter notebook with the code used for this interpolation step.
See section \ref{sec:repro} for details.
We used a mesh with $N=32,768$, and relaxed some parameters compared with the grid-convergence results shown previously, still yielding errors below $1\%$ at all frequencies.
This results in a $12\times$ decrease in the runtime for each case.
The parameters used are shown in Tables \ref{table:quadparams2} and \ref{table:treeparams2}.
Figure \ref{fig:verif_sphere} shows good agreement between the computed and analytical results, 
evidence that \pygbe can accurately represent the mathematical model.

\begin{table}[h]
    \centering
    \caption{\label{table:quadparams2} Verification: Gauss quadrature points; 
    $K$ and $K_{fine}$ are per element; $N_k $ is per element edge (semi-analytical integration). } 
    \begin{tabular}{l l}
    \hline
     distant elements: & $K=4$ \\
     near-singular integrals:   & $ K_{fine}=19$ \\
     singular elements:  & $N_k =9$ \\
    \hline
    \end{tabular}
\end{table}

\begin{table}[h]
    \centering
    \caption{\label{table:treeparams2} Verification: treecode and solver parameters.} 
    \begin{tabular}{l l}
    \hline
    treecode order of expansion: & $P=6$\\
    MAC                                         & $\theta=0.5$\\
    GMRES tolerance                    & $10^{-3}$\\
    \hline
    \end{tabular}
\end{table}

\subsection{LSPR response to bovine serum albumin (BSA)} \label{sec:lspr_response}

Localized Surface Plasmon Resonance (LSPR) biosensors detect a target molecule by monitoring
frequency shifts in the plasmon resonance of metallic nanoparticles, in presence of an analyte \cite{WilletsVandyune2007}.
In this section, we model the response of LSPR biosensors using the expanded capacity of \pygbe.
We consider a spherical silver nanoparticle, and compute the extinction cross-section placing 
bovine serum albumin (BSA) proteins (PDB code: 4FS5, a BSA dimer) in different locations.
We placed two BSA dimers opposite to each other in three configurations ($\pm z$, $\pm y$, and $\pm x$), as shown by figures \ref{fig:display_z} and \ref{fig:display_xy}.
As a point of comparison, experiments by Teichroeb and co-workers \cite{TeichroebETal2008} find a coverage of $2\times 10^{12} \quad \text{molecules}/cm^2$ or $3.3\times 10^{12} \quad \text{molecules}/cm^2$, with a gold sphere 15-nm in diameter. 
In that work, the molecular size reported is 5.5 nm$\times$5.5 nm$\times$9 nm, resulting in a number of attached molecules  between 4 and 6. 
The BSA molecule used in our work corresponds to a dimer, i.e., approximately double the size of that in Teichroeb et al.'s experiment. 
With two BSA dimers in the proximity of the sensor, the volume fractions in the near-by region are comparable.

\subsubsection{Grid-convergence study} \label{sec:bsa_convergence}
We performed a grid-convergence 
analysis of the system sketched in Figure \ref{fig:analyte-sensor}. 
Since we compute the extinction cross-section of the spherical nanoparticle only, we 
set a fixed mesh density for the protein and refined the mesh of the
sphere (meshes of 512, 2048, 8192 and 32768 elements). We found that the protein meshed with two
triangles per ${\AA}^2$ was fine enough for the convergence analysis, resulting in $N_{prot} = 98116$ elements. 

\begin{figure}[h] 
   \centering
   \includegraphics[width=0.45\textwidth]{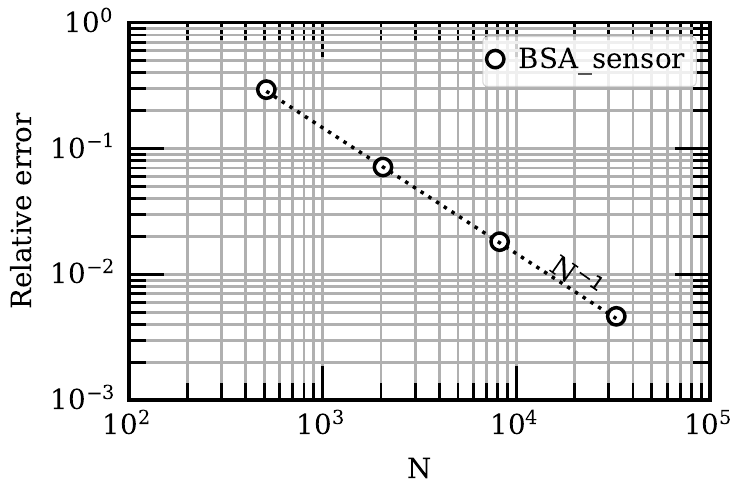} 
   \caption{Grid-convergence study of extinction cross-section of a spherical silver
            nanoparticle with a BSA dimer at $d=1$ nm. Figure, plotting script and auxiliary files available under \textsc{cc-by} \cite{ClementiETal2018c}.}
   \label{fig:error_sphere-bsa}
\end{figure}

We used the same physical conditions as in the grid convergence with an isolated silver nanoparticle, and the same numerical parameters, presented in Tables \ref{table:quadparams1} and \ref{table:treeparams1}.
For the protein dielectric constant, we used $2.7514 + 0.2860i$, obtained from the 
functional relationship provided by Phan, et al.~\cite{PhanETal2013}.
The distance between the sensor and the analyte was $d=1$ nm, 
and the BSA protein was oriented such that its dipole moment was aligned with the $y$-axis. 
To obtain the error estimates shown in Figure \ref{fig:error_sphere-bsa} and Table \ref{table:err_bsa_sensor},
we used the Richardson extrapolated value of extinction cross-section as a reference, 
$C_{ext}= 1778.73$ nm$^2$.

The observed order of convergence is $0.99$, and 
Figure \ref{fig:error_sphere-bsa} shows that the error decays with the number
of boundary elements ($1/N$), which is consistent with our verification 
results in Section \ref{sec:verification}. This provides evidence that the
numerical solutions computed with \pygbe are correctly resolved by the meshes.
The percentage errors for the different meshes are presented in Table \ref{table:err_bsa_sensor}.

\begin{table}
    \centering
    \caption{\label{table:err_bsa_sensor} Estimated percentage error of the BSA-sensor system (Fig.~\ref{fig:analyte-sensor}), with respect to the extrapolated value (using Richardson extrapolation).} 
    \begin{tabular}{c c}
    \hline
    N & \% error \\
    \hline
     $512$ & $29.39$ \\
     $2048$ & $7.13$ \\
     $8192$ & $1.82$ \\
     $32768$ & $0.46$ \\
    \hline
    \end{tabular}
\end{table}

\subsubsection{Resonance frequency shift} \label{sec:bsa_shift}

We computed the LSPR response as a function of the wavelength in the presence 
of the BSA protein. To optimize run-times without compromising accuracy, we used a relaxed
set of parameters, where the protein mesh density was one element per
${\AA}^2$ ($N_{prot}=45140$) and the sphere mesh had $N_{sensor}=32768$ elements. 
These calculations used the same parameters as shown in Tables \ref{table:quadparams2} and \ref{table:treeparams2}.
This parameter choice resulted in a percentage error below 1\%, with respect to the Richardson-extrapolated value.
The run time for each one of these cases was approximately $7.5$ min using one NVIDIA Tesla K40c GPU. 
When two proteins are present, the run time per case is approximately  $15$ min. 

\begin{center}
\begin{figure*}[t] 
   \centering
   \includegraphics[width=0.65\textwidth]{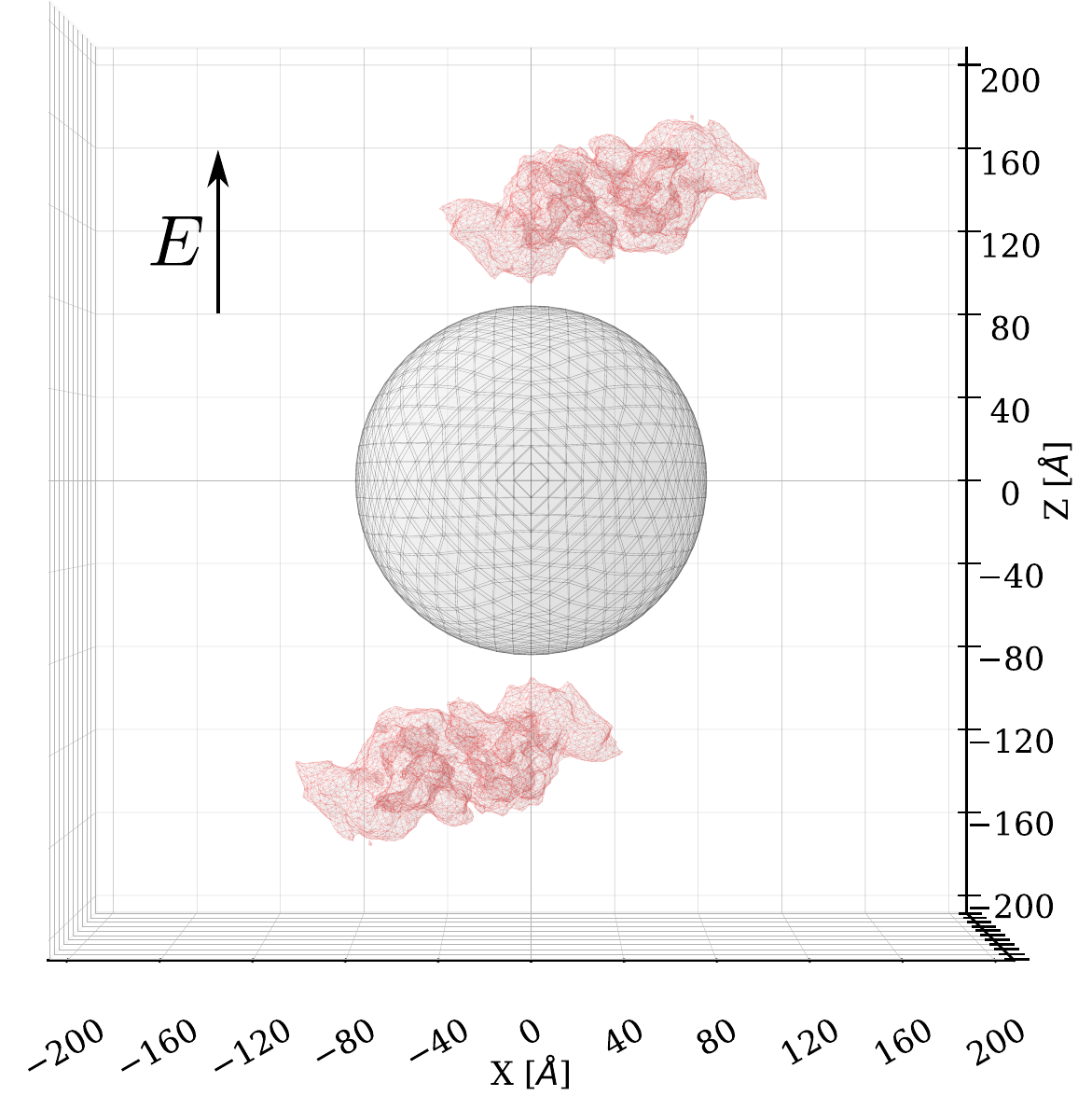} 
   \caption{Sensor protein display: BSA dimers located at $\pm 1$ nm of the 
            nanoparticle in the $z$-direction. Figure, plotting script and auxiliary files available under \textsc{cc-by} \cite{ClementiETal2018e}.}
   \label{fig:display_z}
\end{figure*}
\end{center}

Figure \ref{fig:display_z} shows a visualization of the meshes for these calculations, 
with two BSA proteins placed at a distance $d=1$ nm away from a spherical silver nanoparticle, along the $z$-axis.
The surface-mesh data, plotting scripts and figure are available openly on Figshare, 
in support of the paper's reproducibility \cite{ClementiETal2018e}.
The position of the BSA molecule in the $+z$ axis was the same as in the convergence analysis in 
Section \ref{sec:bsa_convergence}, whereas the BSA in the $-z$ position is a 180$^\circ$ 
solid rotation  about the $y$-axis of the BSA in $+z$.
We performed calculations for wavelengths between $382$ nm and $387$ nm, every $0.25$ nm,
which are around the peak seen in Figure \ref{fig:verif_sphere}.

\begin{figure}[h] 
   \centering
   \includegraphics[width=0.45\textwidth]{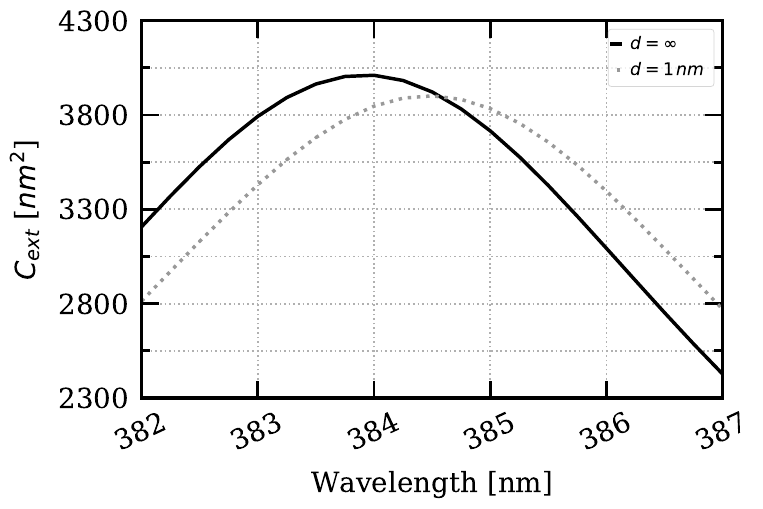} 
   \caption{Extinction cross-section as a function of wavelength for an $8$ nm
            silver sphere immersed in water with two BSA dimers placed 
            $\pm 1$ nm away from the surface in the $z$-direction, and at
            infinity (no protein).}
   \label{fig:2pz_response}
\end{figure}

Figure \ref{fig:2pz_response} shows the variation of the extinction cross-section
with respect to wavelength for the isolated nanoparticle ($d=\infty$) and with
BSA proteins placed $d=1$ nm away. 
The result shows a red-shift ($0.5$ nm) in the resonance frequency due to the
presence of the BSA analytes.

\begin{figure}[t] 
   \centering
   \subfloat{\includegraphics[width=0.45\textwidth]{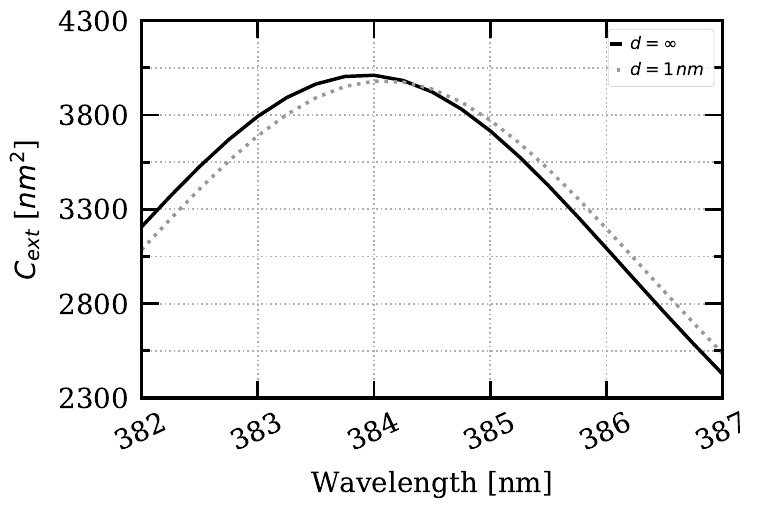}}\\
   \subfloat{\includegraphics[width=0.45\textwidth]{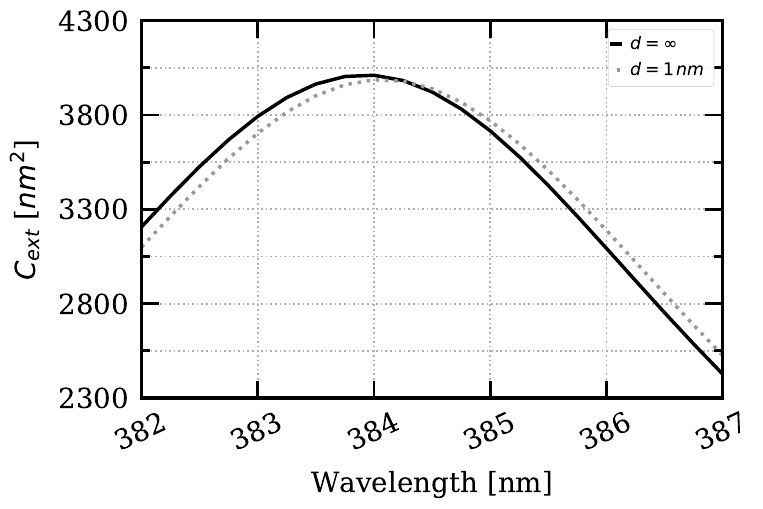}} 
   \caption{Extinction cross-section as a function of wavelength for an $8$-nm
            silver sphere immersed in water with two BSA dimers placed at
            $\pm 1 $ nm away from the surface in the $x$-direction (top) and
             $y$-direction (bottom), and at infinity (no protein).}
   \label{fig:2pxy_response}
\end{figure}

\begin{figure*}
   \centering
   \subfloat{\includegraphics[width=0.65\textwidth]{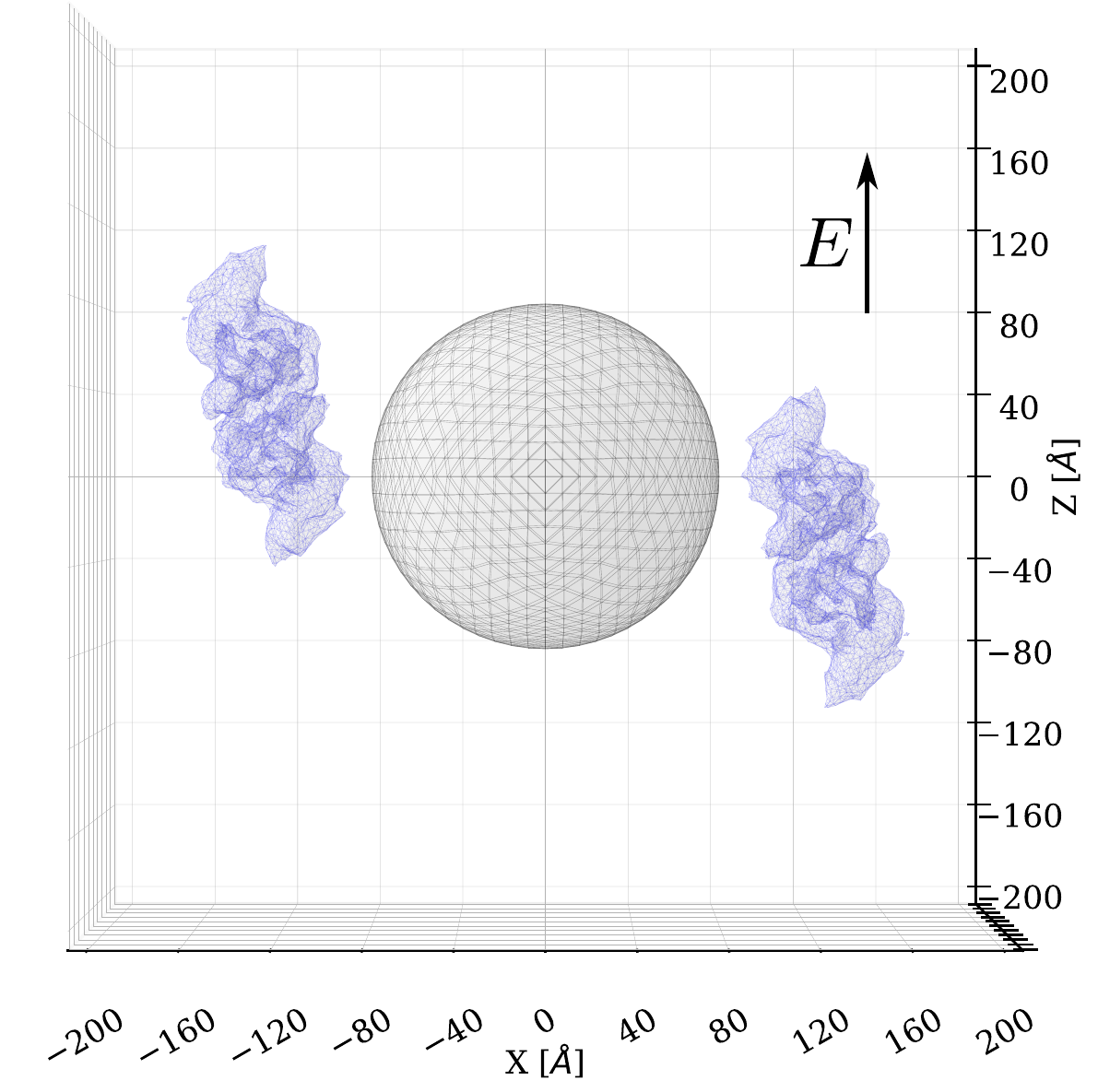}} \\
%
\vspace{-0.5cm}
   \subfloat{\includegraphics[width=0.65\textwidth]{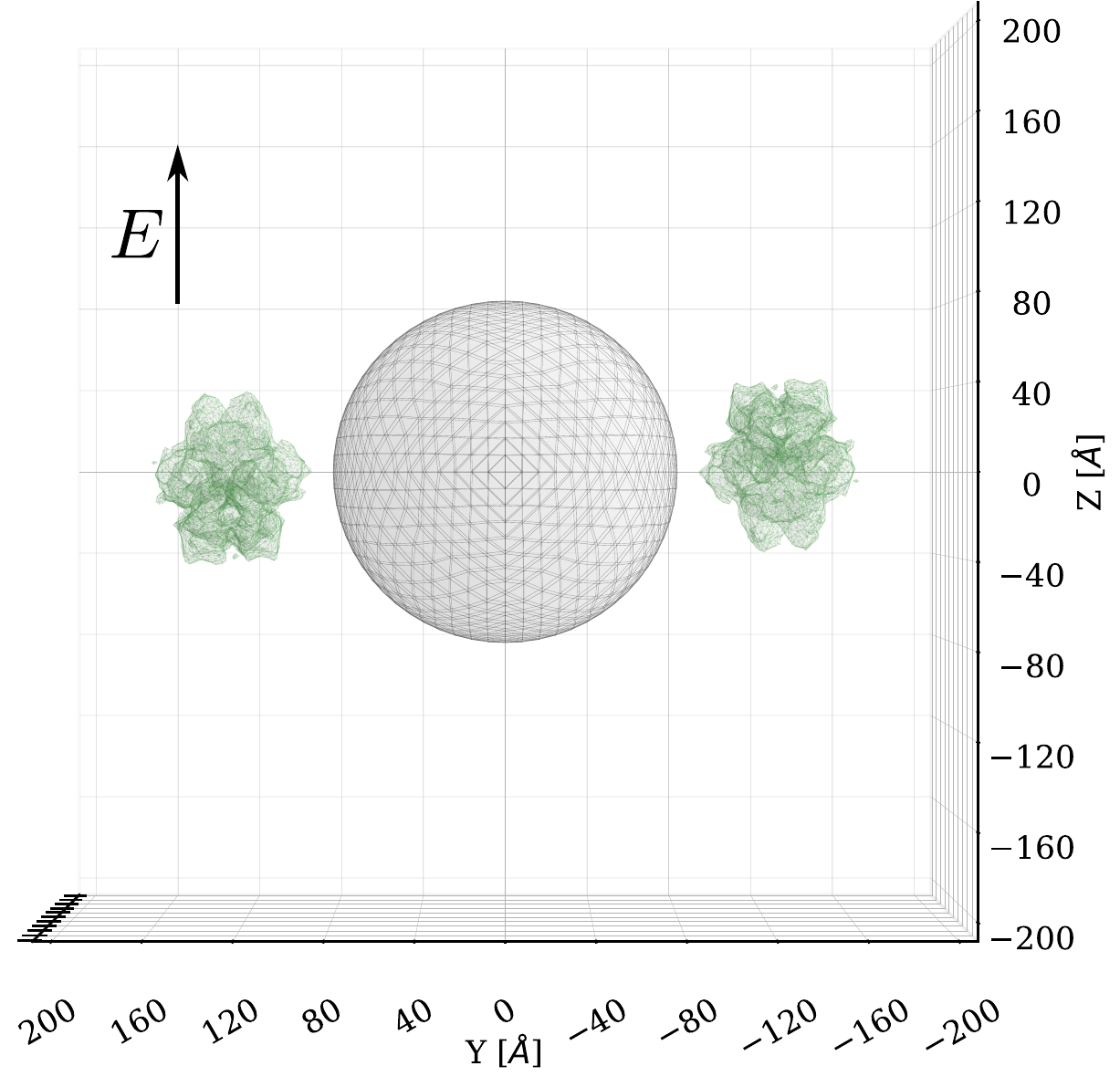}} 
    \caption{Sensor protein display: BSA dimers located at $\pm 1$ nm of the nanoparticle in the
           $x$-direction (top) and $y$-direction (bottom). Figure, plotting script and auxiliary files available under \textsc{cc-by} \cite{ClementiETal2018e}.}
    \label{fig:display_xy}
\end{figure*}

To study the effect of location of the analytes, we re-computed the result placing the 
BSA proteins along the $x$- and $y$-axis, at $\pm 1$ nm, as shown in Figure \ref{fig:display_xy}.
These configurations were obtained via a 90-degree solid rotation of the $z$-configuration (Figure \ref{fig:display_z}) along the $x$- and $y$-axis, respectively.
Figure \ref{fig:2pxy_response} shows the results, in each case.

\subsubsection{Sensitivity calculations} \label{sec:bsa_sensitivity}

\begin{figure}
   \centering
   \includegraphics[width=0.45\textwidth]{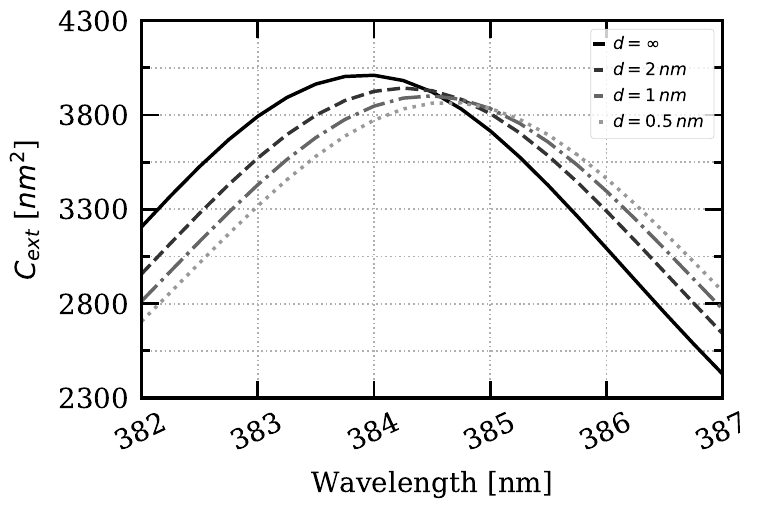} 
   \caption{Extinction cross-section as a function of wavelength for an $8$-nm
            silver sphere immersed in water with two BSA dimers placed at
            $2$, $1$, and $0.5$ nm away from the surface in the 
            $z$-direction, and at infinity (no protein). The test case with $d=0.5$nm is close to the limit where quantum tunneling might happen. Such effects are not captured by our classical model.}
   \label{fig:dist_response}
\end{figure}

The sensitivity of an LSPR biosensor corresponds to the relationship between the size 
of the resonance frequency shift and the number of analytes bound to the sensor (through a ligand).
Experiments show that the distance between the nanoparticle and the analyte 
affects the sensitivity of the sensor, to the point that
targets placed $15$ nm away from the surface are very hard to detect \cite{HaesETal2004}.
This is a critical issue, considering that common ligands (for example, antibodies) can be
larger than $15$ nm. Figure \ref{fig:dist_response} 
shows how the peak varies with the distance at which the analytes ($+z$ and $-z$) are placed.  
In particular, we see a shift of $0.25$ nm when $d=2$ nm to $0.75$ nm when the 
analytes are placed at $d=0.5$ nm. The parameters used in this case remain 
the same as the ones used in Figures \ref{fig:2pz_response} and \ref{fig:2pxy_response} .

\subsection{Reproducibility and data management} \label{sec:repro}

To facilitate the reproducibility and replication of our results, 
we consistently release our research code and data with every publication. \pygbe is openly developed and 
shared under the BSD3-clause license via its repository at \url{https://github.com/barbagroup/pygbe}.

We also release all of the data and scripts needed to run the calculations reported in this work, 
as well as the post-processing scripts to reproduce the figures in this paper. 
All the input files necessary to reproduce the computations are available in one Zenodo data set \cite{ClementiETal2018a}. 
Each problem corresponds to a folder, wherein the user can find geometry files (surface meshes), 
configuration files, parameter files, and when it applies, the protein charges (.pqr).
The scripts and auxiliary files needed to run \pygbe to re-generate every result in the paper are collected in another Zenodo deposit \cite{ClementiETal2018b}.
After execution, the resulting data needed to re-create the figures in the paper will be saved in the running folder and the input files (the first Zenodo set) can at that point be deleted.
(For more details, the reader can consult a \texttt{README} file in the Zenodo archive.)
\emph{Reproducibility packages} to reproduce the figures in the paper are deposited on Figshare, 
including the figures, plotting scripts and Jupyter notebooks that organize and re-create the results \cite{ClementiETal2018c,ClementiETal2018d,ClementiETal2018e,ClementiETal2018f}.

\section{Discussion} \label{sec:discussion}

Extending \pygbe to the LSPR biosensing application required considerable code 
modifications and added functionality. The results presented in the previous section 
offer evidence to build confidence on the suitability of the mathematical model 
and the correctness of the code.
The grid-convergence study with a nanosphere under a constant electric field 
shows a $1/N$ rate of convergence, consistent with convergence results 
in previous work using \pygbe \cite{CooperBardhanBarba2013}.
Further verification of \pygbe's new ability to compute extinction cross-section
of a scatterer in the long-wavelength limit is provided in Figure \ref{fig:verif_sphere}.
The computed extinction cross-section of a silver nanoparticle in a range of frequencies 
is within 1\% of the analytical value, with the numerical parameters chosen.
This level of accuracy is likely sufficient, given that experimental uncertainty in 
the values of the dielectric constant for silver is in the order of 1\%, also \cite{JohnsonChristy1972}.


Figure \ref{fig:2pz_response} shows a red shift of the plasmon resonance frequency peak in presence of the BSA proteins.
Experimental observations of Tang, et al.~\cite{TangETal2010} with silver nanoparticles of approximately 17 nm in diameter and BSA proteins in solution revealed a red shift upon adding the proteins. 
Similar to the effect we see with our model, they observed as well a decrement of 
the peak amplitude.
Moreover, recent experiments \cite{PuETal2018} report a resonance frequency for a silver nanoparticle in the presence of BSA proteins of between 380 and 400 nm, which is consistent with our results.
Other experiments \cite{RaphaelETal2013} also report a red shift in the resonance frequency in the presence of (different) proteins.
Our boundary element method approach using electrostatic
approximation is thus able to capture the characteristic resonance-frequency 
shift of LSPR biosensors.

With the electric field aligned in the $z$-direction, placing the proteins at a distance
in the $x$ or $y$ directions from the nanoparticle shows a negligible shift in the 
resonance peak: the shifts in Figure \ref{fig:2pxy_response} 
are smaller than the resolution between wavelengths ($< 0.25$ nm).
This finding is consistent with the free electrons oscillating in the $z$ direction
under a $z$-polarized electric field, and not in the $x$ and $y$ directions
(see Figure \ref{fig:lspr}). 
The analytes have a marked effect when placed in the $z$ direction, where
they can interfere with the free oscillating electrons. 

Figure \ref{fig:dist_response} shows how the  shift in resonance frequency varies 
with the distance between the sensor and the analyte. As expected, the shift decays 
as the BSA moves away from the sensor, to the point that if the BSA proteins are placed
$d=2$ nm away, the shift is only $0.25$ nm. This result shows the potential of \pygbe 
and the electrostatic approach to study biosensor sensitivity with distance.
Note that possible quantum effects (e.g., tunneling) at $d=0.5$nm are ignored with our classical model.
Even if this distance could be close to or in the quantum regime, evidence that classical theory is valid at this distance in similar systems has been reported 
\cite{SavageETal2012, EstebanETal2012}.  

Even though there is evidence that techniques such as Plasmon Enhanced Raman Scattering are capable of detecting all the way to single molecules \cite{ZhangZhangETal2013}, 
as far as we know, there is no evidence of purely LSPR approaches that can sense such low concentration of analytes.
These computational studies can shine light on  potential improvements that would enhance sensitivity of LSPR biosensors, for example, by using smaller ligands. 

We are not aware of other LSPR simulations where the molecular details of the analyte are considered, however, similar calculations could be performed with other software. 
For example, BEM++ \cite{SmigajETal2015} also models the system as a set of boundary integral equations, discretized in flat triangular panels. 
This software uses the Galerkin approach and algorithmic acceleration via hierarchical matrices, which is slower and less memory efficient than the treecode and limits the accessible problem sizes.
The Matlab toolbox MNPBEM \cite{HohenesterTrugler2012} is another alternative software designed to simulate scattering of metallic nanoparticles.
Its BEM implementation is similar to \pygbe as it uses a centroid collocation scheme on flat triangular panels, but differs in the algorithmic acceleration technique, which is also based on hierarchical matrices rather than a treecode. 
This results in higher memory usage compared to our code, making it harder to simulate large analytes in detail.
Commercial finite-element or finite-difference solvers could also be used in this application, for example, COMSOL. 
These volumetric approaches, however, struggle to correctly impose the zero boundary condition at infinity, which is exactly met for a BEM formulation.

\section{Conclusion}

In this work, we combined the implicit-solvent model of electrostatics interactions in \pygbe 
with a long-wavelength representation of LSPR response in nanoparticles. 
We extended \pygbe to work with complex-valued quantities, and added functionality to 
include an imposed electric field and compute relevant quantities 
(dipole moment, extinction cross-section). 
Previous work with \pygbe showed its suitability for computing 
biomolecular electrostatics considering solvent-filled cavities and Stern layers \cite{CooperBardhanBarba2013}, 
and for protein-surface electrostatic interactions \cite{CooperBarba2016}.
This latest extension can offer a valuable computational approach to study nanoplasnomics and aid in the design of LSPR biosensors. 
Thanks to algorithmic acceleration with a treecode, and hardware acceleration with GPUs, \pygbe is able to compute problems with half a million elements, or more, which is required to represent the molecular surface accurately.

\begin{acknowledgments}

CDC acknowledges the financial support from CONICYT through projects FONDECYT Iniciaci\'on 11160768 and Basal FB0821.
\end{acknowledgments}

\bibliography{compbio,bem,scicomp,fastmethods,biosensors} 

\end{document}